\documentclass[12pt]{iopart}

\usepackage{cite}

\usepackage[english]{babel}
\usepackage{latexsym}
\usepackage{graphics,psfrag}
\usepackage{epsfig}
\usepackage[utf8]{inputenc}
\usepackage[T1]{fontenc}
\usepackage[usenames,dvipsnames]{color}
\def\be{\begin{equation}}
\def\ee{\end{equation}}
\def\bea{\begin{eqnarray}}
\def\eea{\end{eqnarray}}
\def\bi{\begin{itemize}}
\def\ei{\end{itemize}}
\def\bin{\begin{enumerate}}
\def\ein{\end{enumerate}}

\begin{document}

\title{Dynamics of cold bosons in optical lattices: \\ Effects of higher Bloch bands}


\author{Mateusz \L\k{a}cki}
\address{
Instytut Fizyki imienia Mariana Smoluchowskiego, 
Uniwersytet Jagiello\'nski, ul.~Reymonta 4, PL-30-059 Krak\'ow, Poland}

\author{Dominique Delande} 

\address{Laboratoire Kastler-Brossel, UPMC-Paris 6, ENS, CNRS;  
4 Place Jussieu, F-75005 Paris, France}
 
\author{Jakub Zakrzewski} 

\address{
Instytut Fizyki imienia Mariana Smoluchowskiego, 
Uniwersytet Jagiello\'nski, ul.~Reymonta 4, PL-30-059 Krak\'ow, Poland}

\address{
Mark Kac Complex Systems Research Center, 
Uniwersytet Jagiello\'nski, ul.~Reymonta 4, PL-30-059 Krak\'ow, Poland}

\date{\today}

\begin{abstract}

The extended effective multiorbital Bose-Hubbard-type Hamiltonian which takes into account higher Bloch bands, 
is discussed for boson systems in optical lattices, with emphasis on dynamical properties, in relation with current experiments. 
It is shown that the renormalization of Hamiltonian parameters
depends on the dimension of the problem studied. Therefore, mean field phase diagrams do not scale with the coordination number of
the lattice. The effect of Hamiltonian parameters renormalization on the dynamics in reduced one-dimensional optical lattice
potential is analyzed. We study both the quasi-adiabatic quench through the superfluid-Mott insulator 
transition and the 
absorption spectroscopy, that is energy absorption rate when the lattice depth is periodically modulated.   
\end{abstract}


\maketitle

\section{Introduction}
Ultracold bosonic atom gases in optical lattices have been an ultrahot research area in recent years 
(for recent reviews and an extensive reference list see \cite{Lewenstein2007,Yukalov2009, Lewenstein2012}). 
They provide means to create and control experimental systems mimicking different condensed matter physics models \cite{Greiner2002,Paredes2004,Bloch2005}. 
The interest
has been stimulated in part by the fact that there exists \cite{Jaksch1998} an accurate mapping of continuous 
Hamiltonian to a~Hamiltonian on a~lattice --- the Bose-Hubbard (BH) Hamiltonian,
originally formulated by Gersch and Knollman \cite{Gersch1963}.  The lattice models significantly ease
the analytical \cite{Fisher1989,Oosten2001,Freericks1996,Damski2006} and numerical analysis, though
promising new ideas were recently proposed  that can enable the analysis in continuous variables \cite{Verstraete2010} 
beyond the mean field level \cite{Rokhsar1991,Zakrzewski2005}. 

Bosons in one  dimensional (1D) optical lattices have also been the area of extensive experimental research
\cite{Stoferle2004,Kinoshita2004,Fallani2007,Gemelke2009,Bloch2012,Trotzky2012}. The corresponding 1D BH Hamiltonian can be
effectively addressed numerically by Density Matrix Renormalization Group (DMRG) and related techniques \cite{Schollwock2011}. These
techniques have broad applications, in particular enable simulations targeting real-life many-body systems \cite{White2004}, with
controllable error and no systematic errors due to unsounded assumptions. Their success relies on area laws that control the growth
of entanglement \cite{Eisert2010}. The entanglement is used as a~'small parameter'\cite{Vidal2003,Vidal2004} that makes it 
possible the construction of  an efficient variational set - the so called Matrix Product States (MPS). Several numerical
investigations of experimental and `close to experimental' systems have been performed \cite{Kollath2006,Clark2006,Zakrzewski2009}.
They focused largely on two aspects: a~quench through a~phase transition and the simulation of absorption spectroscopy
\cite{Stoferle2004}. The phase transition from the superfluid to the Mott insulator phase occurs when the lattice depth is
increased beyond a~critical value\cite{Greiner2002,Cucchietti2007}. The adiabaticity of this process has been addressed in
\cite{Zakrzewski2005,Zakrzewski2009,Eurich2011}. In the second example, the energy absorption rate is analyzed as a~function of the 
frequency of modulation of the lattice depth  \cite{Kollath2006,Huo2011,Lacki2012}. The locations, number of peaks in the spectrum
and their heights are related to the state of the gas: either superfluid or insulating.

The derivation of the BH Hamiltonian assumes the restriction of physics to the lowest Bloch band of the optical lattice. 
This assumption is reasonable, as the energy gap between the first and second Bloch bands is in most cases around
10 times larger than the energy scale in the discrete model. It is also much larger than the thermal energy scale 
$k_B T$ providing additional argument for a zero temperature analysis.
An early analysis \cite{Oosten2003} suggested that  higher bands may be included by an appropriate modification of
BH Hamiltonian parameters for large occupation numbers. The effects due to  higher Bloch bands have been also studied for lower
densities \cite{Johnson2009,Larson2009,Dutta2011,Sakmann2011} in relation to the discovery that higher bands affect the
superfluid-insulator transition in Bose-Fermi mixtures \cite{Luhmann2012,Mering2011}. This gave explanation for  experimental
observations of the shift of the Superfluid (SF) -- Mott insulator (MI) phase transition  \cite{Ospelkaus2006} which could not be
explained by single-band approaches. In \cite{Larson2009}, the first excited band is included in a~two-flavour model;
effects of higher bands could also be built in via an effective three-body interaction in the lowest band \cite{Johnson2009,
Dutta2011}. 

Recently, another approach for  3D optical lattices  
has been proposed \cite{Will2010,Bissbort2011,Luhmann2012} (which somehow
resemble in spirit  \cite{Oosten2003} while being used for moderate atomic densities). In this approach, the higher bands are
included in the on-site Hamiltonian, which is then diagonalized  and many-body ground states of this problem for different total
number of particles are used as a~local Hilbert space, replacing the usual Fock basis $|n\rangle.$ 
The multiband lattice
Hamiltonian is then expressed in this new basis, yielding an effective single-band model with occupation dependent parameters ---
renormalized values of the initial Hamiltonian parameters for the lowest Bloch band. Interestingly, the change of BH Hamiltonian
parameters is large in the MI regime, where the energy gap is also large, contrary to a~naive intuition.

In this paper, we consider the derivation of the BH model's effective coupling constants, taking a~closer look at the underlying,
quite challenging, numerical problem --- the accurate diagonalization of the onsite Hamiltonian.  
We study how the dimensionality of the optical lattice affects the renormalization scheme. The dependence on dimension implies
that the mean field diagrams of the system no longer depend solely on the coordination number of the lattice.
We then take a~look at the effects of the renormalization of coupling constants  on the dynamics in a~1D optical lattice: both
quench through the SF-MI transition and the absorption spectroscopy are analyzed. We find that renormalization of atom-atom
interactions significantly shifts and sometimes modifies the absorption peaks. It also appears to have a~serious effect on
adiabaticity predictions for the SF-MI phase transition.

\section{Tight-binding descriptions of an ultracold boson gas in an optical lattice}

The second quantization Hamiltonian for a~dilute gas of interacting bosonic atoms 
in the optical lattice potential $V(\vec r)$  and external trapping potential $V_e(\vec r)$ is of the form:
\begin{eqnarray}
 H&= &\int \textrm{d}^3\vec r\ \Psi^\dagger(\vec r)\left(-\frac{\hbar^2}{2 m } \nabla^2+V(\vec r)+V_e(\vec r)\right)\Psi(\vec r)  \nonumber\\
&+&\frac{1}{2}\int\textrm{d}^3\vec r \textrm{d}^3\vec r'  \Psi^\dagger(\vec r) \Psi^\dagger(\vec r')V_{\mathrm{int}}(\vec r, \vec r') \Psi(\vec r) \Psi(\vec r'), 
\label{eqn:SecQ}
\end{eqnarray}
where $V_{\mathrm{int}}(\vec r,\vec r')$ is  an isotropic short-range pseudopotential modelling s-wave  interactions \cite{Bloch2008}  
\begin{equation} 
V_{\mathrm{int}}(\vec r,\vec r')=\frac{4\pi \hbar^2 a_s}{m}\delta(\vec r-\vec r')\frac{\partial}{\partial |\vec r-\vec r'| }|\vec r-\vec r'| 
\label{eqn:Vint}
\end{equation}
with $a_s$ being the scattering length.

The Hamiltonian admits a~natural energy scale, set by the recoil energy $E_R = \frac{\hbar^2 k^2}{2m}, $ where
$k=\frac{2\pi}{\lambda}$ and $\lambda$ denotes the optical lattice wavelength. We assume this energy unit from now on. 
The optical lattice  potential is typically: $V(x,y,z)/E_R= s_x \cos^2(k x) 
+ s_y \cos^2(k y)  +s_z \cos^2(k z).$ If $s:=s_x \ll s_y=s_z=:s_\perp,$ then a~classical setup 
for a~1D optical lattice is obtained
\cite{Stoferle2004}. Tunneling in $y,z$ directions is highly suppressed and the system may be considered 
as a~series of independent
1D tubes  along the $x$ direction. Similarly when $s:=s_x=s_y \ll s_z=:s_\perp$ a~2D optical lattice  is obtained. When
$s_x=s_y=s_z=:s$ the potential corresponds to a~3D lattice.
 Let us emphasize that all calculations of renormalized parameters presented below are truly three-dimensional
ones, the labels ``1D'' or ``2D'' just refering to situations where  $s$ and $s_\perp$  are very different. 

Jaksch and Zoller in their seminal paper \cite{Jaksch1998} introduced a~mapping of  (\ref{eqn:SecQ}) onto the BH Hamiltonian (here
$V_i$ is the local energy shift due to $V_e:$  $V_i=V_e(\vec r_i)$):
\begin{eqnarray}
H_{BH}&=&-J \sum\limits_{\langle i j \rangle} \hat a_i \hat a_j^\dagger +h.c. + \frac{U}{2} \sum\limits_i \hat n_i(\hat n_i-1) -
\sum\limits_i\hat n_i (\mu - V_i).
\label{eqn:HamBH}
\end{eqnarray}
The field operator is expanded in  the set of the lowest band Wannier functions of the lattice: 
$\psi(\vec r)=\sum\limits w_i^0(\vec r) a_i,$ ---
here $w_i^\alpha$ denotes the (real valued) Wannier function localized on site $i$ of the lattice for the $\alpha$'th Bloch
band (we shall denote sites by roman subscripts and bands by greek superscripts). The parameters $U$ and $J$ are expressed by the
appropriate integrals of Wannier functions: $J=-\int w_i^0( \vec r) \left(-\frac{\hbar^2}{2m}\nabla^2 + V(\vec r)
\right)w_{j}^0(\vec r)\ d\vec r$ (where $i$ and $j$ are neighbouring sites) and $U= \frac{4\pi \hbar^2 a_s}{m} \int w_i^0(\vec r)^4 d\vec
r.$ 

However, it turns out that such a single band approximation is often insufficient for realistic values
of the parameters, and that contributions of higher bands cannot be neglected. This problem has been studied
in the literature in various conditions and using various methods. Of special interest is the situation where the scattering
length $a_s$ is large (near a Feshbach resonance), where the optical lattice strongly modifies the effective atom-atom interaction,
see e.g.~\cite{Fedichev2004,Dickerscheid2005,Gubbels2006,Diener2006,Wouters2006,Buechler2010,Cui2010,VonStecher2011}.
This is not the situation realized for the $^{87}$Rb atom in zero magnetic field used in the Florence experiment~\cite{Fallani2007,Luhmann2012}
where $a_s=5.2\mathrm{nm}$, much shorter than the lattice spacing $\lambda/2=377\mathrm{nm}.$ 

In  this paper, we will consider only situations where the atomic wavenumber $k$ -- being evaluated either in the lowest band or in the excited bands
included in the calculation -- is such that $ka_s \ll 1,$ so that only s-wave low energy interatomic scattering is relevant.
In practice, this puts a limit on the number of bands used $B<20,$ beyond which the model is not a good approximation
of real world.

It is well known that, in dimension higher than 1, contact interactions cannot be modelled by a $\delta$-potential, but require a
specific regularization to avoid artificial divergences. Mathematically, a self-adjoint extension of the original Hamiltonian
is needed~\cite{Albeverio1988}. It boils down to a dimension-dependent regularization of the $\delta$-potential~\cite{Wodkiewicz1991},
which, in 3D, is the so-called Fermi pseudo-potential~\cite{Fermi1936} used in Eq.~(\ref{eqn:Vint}). Even with the correct pseudo-potential,
one must be careful when expanding over an infinite set of square integrable smooth basis functions (such as the Wannier functions used below)
without renormalization of the interaction strength, because it may lead to incorrect results, such as diverging perturbative expansions,
see \cite{Busch1998} for the specific example of two interacting particles in an harmonic trap. Numerical diagonalization,
used in the following, leads to less severe problems as discussed in~\cite{Rontani2008}. 
We will use a rather small number of bands (up to $B$=15), so that the highest atomic wavenumbers effectively included in the calculation
are still rather small and the divergence of the Green function at short distance is just a small perturbation. In other words,
although the method we use leads in principle to divergences when $B\to \infty,$ these divergences manifest themselves
only beyond the largest $B$ used in our calculations.
Note also that the single band approximation described above is free of this problem and pure $\delta$ interactions can be used in these cases.

Without restriction to the lowest Bloch band, the expansion in the full Wannier functions basis set would give a~multiband variant
of the Bose-Hubbard Hamiltonian:  
\begin{eqnarray} 
 H&=&-\sum\limits_{\alpha, i,j}J^{\alpha\alpha}_{ij}   (\hat a^{\alpha}_{i})^\dagger \hat a^{\alpha}_{j} + h.c. +
\sum\limits_{\stackrel{\alpha\ldots\delta}{i\ldots l}}U^{\alpha\beta\gamma\delta}_{ijkl} (\hat a^{\alpha}_i)^\dagger (\hat
a^{\beta}_{j})^\dagger \hat a^{\gamma}_k \hat a^{\delta}_{l}\nonumber \\
 & & +  \sum\limits_{\alpha,i} (E_\alpha +V_i - \mu) (\hat a^{\alpha}_{i})^\dagger \hat a^{\alpha}_{i} 
  \label{eqn:HamFull}
 \end{eqnarray}
with 
\begin{equation}
U^{\alpha\beta\gamma\delta}_{ijkl}= \frac{4\pi \hbar^2 a_s}{m} \int d\vec r\ w^{\alpha}_{i} (\vec r) w^{{\beta}}_{j} (\vec r)w^{{\gamma}}_{k} (\vec
r)w^{{\delta}}_{l} (\vec r ),
\label{eqn:Udef}
\end{equation} 
and 
$$J^{\alpha\alpha}_{ij} =-\int  d\vec r\ w^{\alpha}_{i}(\vec r) \left(-\frac{\hbar^2}{2m} \nabla^2+V(\vec r)\right)w^{\alpha}_{j}(\vec r).$$
Note that, because Wannier functions are smooth, the potential (\ref{eqn:Vint} may be replaced by a contact Fermi potential
\begin{equation} 
V_{\mathrm{c}}(\vec r,\vec r')=\frac{4\pi \hbar^2 a_s}{m}\delta(\vec r-\vec r')
\label{eqn:Vcon}
\end{equation}
 in the integral~(\ref{eqn:Udef}).

By construction $J_{ij}^{\alpha\beta}=0$ for  $\alpha\neq \beta.$ 
For sufficiently deep optical lattices (typically $s>3$), it is enough to restrict hopping to nearest-neighbor sites,
as tunneling amplitudes are exponentially damped with the hopping distance (for shallow lattice next nearest neighbours
hopping may be necessary - see \cite{Trotzky2012}).
  
A 3D Wannier function being a~product of 1D  Wannier functions,
 the 3D integral in $U^{\alpha\beta\gamma\delta}_{ijkl}$ is a~product of 3 integrals over each coordinate. The interaction
parameters differ for each direction, as Wannier functions depend on the lattice depth, which may be different in each direction.

The Hamiltonian (\ref{eqn:HamFull}) is difficult to use in practice, even in 1D
systems, because the onsite dimension $d$ of the $n$-particle problem restricted to the lowest $B$ bands is $d={B^3 + n-1 \choose
n},$ (the onsite problem is genuinely 3D even for a~quasi-1D models) and moreover the numerical complexity of the best 1D algorithms
scales with $d$  at least as $O(d^3).$ The complexity of more sophisticated approaches (such as MERA, PEPS
\cite{Evenbly2009,Verstraete2006,Verstraete2004,Vidal2007}) is several orders higher. 

Thus for computational purposes, one must restrict the local Hilbert space.
Assuming  that  interactions are on-site only, i.e.   $U^{\alpha\beta\gamma\delta}_{ijkl}\ne0$ for $i=j=k=l$ together with
considering the lowest Bloch band only ($\alpha=\beta=\gamma=\delta=0$) \cite{Jaksch1998} leads directly to the~Bose-Hubbard
Hamiltonian, Eq.~(\ref{eqn:HamBH}), provided we chose the zero of the energy axis at $E_0.$ 

A more sophisticated approach is discussed in \cite{Bissbort2011,Luhmann2012}.  
The on-site Hamiltonian, restriction of
(\ref{eqn:HamFull}) to a single site:
\begin{equation}
\label{eq:onsiteham}
\mathcal{H}_{loc}= H_E + H_U = \sum\limits_\alpha E_\alpha \hat n_\alpha + \sum\limits_{\alpha\beta\gamma\delta}
U^{\alpha\beta\gamma\delta} \hat a^\dagger_\alpha  \hat a^\dagger_\beta  \hat a_\gamma  \hat a_\delta.
\end{equation}
(with $\hat n_\alpha=\hat a^\dagger_\alpha \hat a_\alpha$) can be diagonalized to yield a~space of $n$ particle ground
states. The eigenenergies $\epsilon_0^n$ of the 
on-site $n$ particle ground states $|\psi_0^n\rangle,$  are the starting point in determining new values of $U$ parameters in the
effective Hamiltonian. To define renormalized values of $U,$ the energy $\epsilon_0^n$ 
has to be decomposed into the interaction energy [which in case of the BH model, is just $\frac{U}{2}  n(n-1)$] and
a~single-particle energy (which in the BH case  shifts $\mu$ by the lowest Bloch band energy). The most natural way to
define the interaction
energy would be to use:

\begin{equation}
 \frac{U_n}{2} n(n-1)   = \langle \psi_0^n |\sum\limits_{\alpha\beta\gamma\delta}
U^{\alpha\beta\gamma\delta} \hat a^\dagger_\alpha  \hat a^\dagger_\beta  \hat a_\gamma  \hat a_\delta
 |\psi_0^n\rangle. 
\label{eqn:TheBest}
\end{equation}

Unfortunately, $U_n$ cannot be defined in such a~way if we request the Hamiltonian to have a form resembling
Eq.~(\ref{eqn:HamBH}). That is because the single particle energy is no longer a~linear function of $n.$ 
This can be circumvented by defining $U_n$ via:

\begin{equation}
\epsilon_0^n   = \frac{U_n}{2} n(n-1) + n E_0 .
\label{eqn:Un}
\end{equation}
This definition makes Hamiltonians (\ref{eqn:HamBH}) and (\ref{eqn:HamEff}) similar. But $U_n$ is no longer the interaction energy,
it also contains contributions of higher Bloch bands population to the single-particle energy. From now on we use definition
(\ref{eqn:Un}). Note that $U_n$ depends nontrivially on the geometry of the lattice.

The second stage is to reintroduce inter-site couplings. Even if only one band is
taken into account (so that $U_n$ is simply $U$), the inter-site interaction $U^{0000}_{iiij}$ induces an effective
coupling, which is proportional to the density, that is a term $U^{0000}_{iiij}a_ia_j^{\dagger}(n_i+n_j-1)+h.c.$
in the effective Hamiltonian (called bond-charge term in 
\cite{Luhmann2012,Sowinski2012a}). 
Similarly to the original tunneling term, it is important only between nearest neighbors as soon as $s$ is larger than unity.
In the low-density regime (typically $n<7$), this contribution leads to an
increase  (since  $U^{0000}_{iiij}<0$ for $s>1$) 
of the tunneling amplitude $J\to J-U^{0000}_{iiij}(n_i+n_j-1).$ The correction is at most of the order of the raw
tunneling. 
When higher bands are taken into account, the modification of $|\psi_0^n\rangle$ induces a renormalization
of the standard tunneling term (as well as of the bond-charge term)
which becomes also dependent on the occupation numbers of sites between which tunneling occurs. 

The effective multiorbital (EMO) Hamiltonian finally becomes:

\begin{equation}
H^{EMO}= - \sum_{\langle i,j \rangle} (a_i a_j^\dagger \sum_{n_i,n_j}J_{n_i,n_j} P^i_{n_i}P^j_{n_j}+ h.c.)   + 
\sum_{n,i}\frac{U_n}{2}n(n-1) P^i_{n},
\label{eqn:HamEff}
\end{equation}
where $P^i_{n} = |i ,n \rangle \langle i , n |.$

This Hamiltonian will allow us to study the influence of higher bands on the dynamics later. 
First we discuss 
the accurate numerical determination of the $U$ and $J$ parameters that, in itself, is a~challenging problem, 
giving an insight into the
physics involved.

\subsection{Solving the onsite problem}

The single site problem is equivalent to finding the $n$ particle ground state
of the Hamiltonian (\ref{eq:onsiteham}).
If the lowest $B$ Bloch bands are included (typical values of $B:$   4 --- \cite{Will2010}, 9 --- \cite{Luhmann2012}), the problem
quickly becomes too involved computationally to be exactly diagonalized, and truncation  of the basis has to be performed.

For small particle numbers, the problem is dominated by the $H_E$ term. The direct transition from the lowest to the first excited
band is forbidden by symmetry consideration so the relevant energy scale is $\Delta=E_2-E_0$. 
Promoting a particle from the lowest
to the second excited band, via interaction term is proportional to $n^{3/2}U^{2000}$ where $U^{2000}$ is the corresponding
interaction integral. This yields for the validity of the perturbative approach the condition $n^{3/2}U^{2000}\ll \Delta$. For typical
parameters corresponding to Rb scattering length and $s$ of the order of 30, $\Delta/U^{2000}\approx 60$  yielding the limiting
value of $n\ll 15$.
 
For small occupation numbers, a perturbative approach seems justified. Perturbation theory enables us to
estimate the impact each excited vector $|\psi_p\rangle$ has on the ground state energy. Let $|\psi_0\rangle$ be an $n$-boson ground (Fock) state of $H_E$. The larger the matrix element 
$|\langle \psi_0 | \mathcal{H}_{loc}| \psi_p\rangle |^2$ and the smaller the energy $\langle \psi_p | \mathcal{H}_{loc} | \psi_p
\rangle $ the larger is the impact. 
The perturbative scheme provides a~hint on how to choose an ``optimal''
subset of the basis in which the full problem could be diagonalized. 
Since the exact diagonalization in the variational basis is the last step, 
we do not follow perturbation theory exactly, but 
just use it to choose a~close to optimal basis, not to calculate the energy correction. 
Details of basis generation and variational space sampling are given in the  Appendix. A more traditional method
\cite{Luhmann2012} is to choose a~subset according to least energy principle --- with minimal $\langle \psi_p |
\mathcal{H}_{loc} | \psi_p \rangle.$

We have performed a~detailed analysis comparing both methods for $s=s_\perp=34.8,$ a~strongly coupled case (3D optical lattice,
$2a_s/\lambda=0.014, \lambda=754 \textrm{ nm}$ --  parameters taken as typical values from \cite{Luhmann2012}). We choose
a~system with $n=2-5$ particles and $40000$ basis vectors according to both least-energy (as in  \cite{Luhmann2012}) and the
perturbative method (39900 vectors are generated within the first order, and 100 within the second order perturbative scheme).

\begin{figure}[ht]
\centering
\includegraphics[clip,width= 12cm]{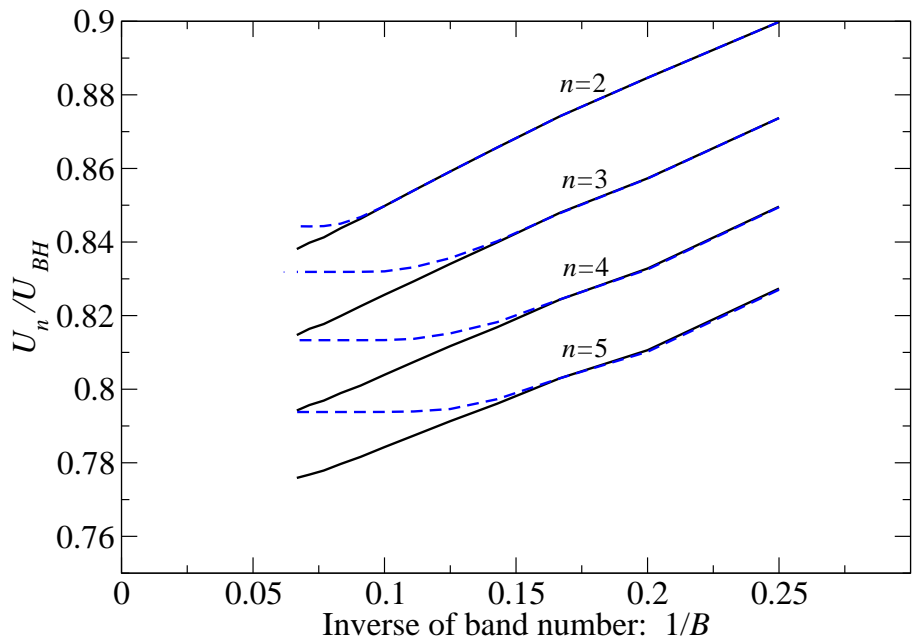}
\caption{Comparison of the effective on-site interaction strength $U_n$ obtained using diagonalization of the
on-site Hamiltonian on two different basis
sets with the same size equal to 40000. If basis vectors are chosen according to their energy (dashed lines), 
false saturation effects appear.
Estimating the influence by a perturbative-like scheme (solid lines) does not seem to suffer from saturation effects. The
3D case is considered: $s=s_\perp=34.8.$ }
\label{fig:comp}
\end{figure}

The ground state energy -- obtained from numerical diagonalization of the on-site Hamiltonian in a restricted subset -- 
versus the number $B$  of bands included is shown in  Figure~\ref{fig:comp}, for various $n$.
The least energy method clearly gives the false impression of saturation of results 
if $B\approx7-9$ bands are included. The false
saturation occurs because the least-energy method does not evaluate  $|\langle \psi_0 | \mathcal{H}_{loc}| \psi_p\rangle |^2$  
and therefore fills the variational basis with low-energy irrelevant vectors with vanishingly small matrix elements. 
The perturbative-like approach does not show similar saturation effects and, moreover, suggests that linear extrapolation of the
results may be performed. We find that the best compromise between computational effort and accuracy is to perform extrapolation of
results as a~function of $1/B$ by means of the ansatz $U_n(B) = U^\infty_n+ c_0/B.$ The same extrapolation scheme can be used for
the $J$ parameters (but leads to less drastic modification of the results).
As mentioned above, numerical diagonalization in the set of Wannier functions might lead to
unphysical divergences as $B\to \infty$ because of the subtle properties of contact interactions in 3D. One could expect a $1/B$
divergence for large $B.$ Figure~\ref{fig:comp} does not show any indication of such a divergence,
which could be visible for larger $B.$

\begin{figure}[ht]
\centering
\includegraphics[clip,width= 12cm]{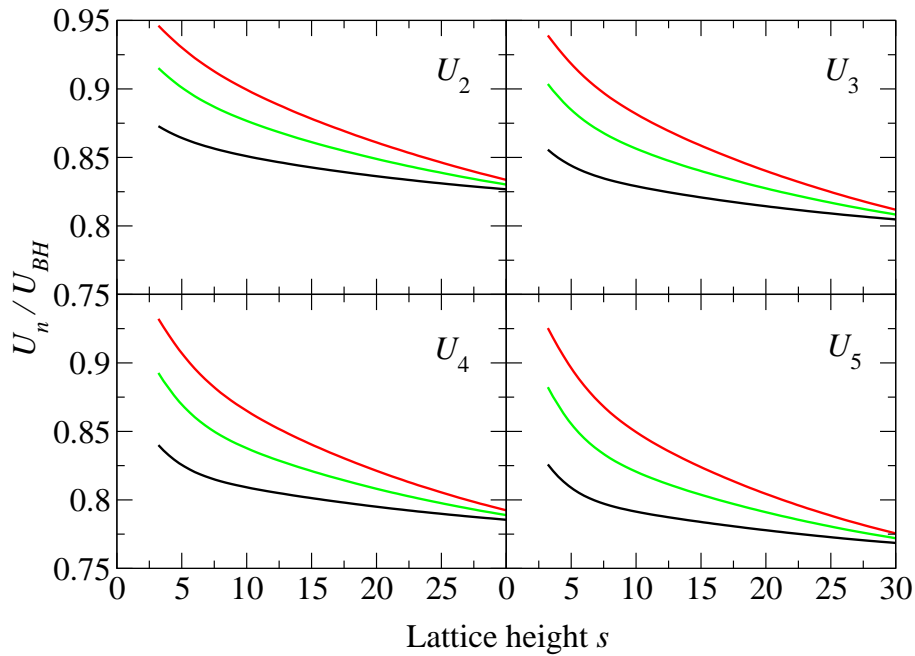}
\caption{Renormalized interaction parameters $U_n$ vs. strength of the optical lattice, for various dimensionalities:
black, green, and red lines corresponding to 1D, 2D, 3D, respectively. 
Interaction and lattice  parameters: $2a_s/\lambda=0.014, \lambda=754
\ \textrm{nm}.$ The transverse lattice height is $s_\perp=34.8.$ The curves meet at 
$s=s_\perp.$} 
\label{fig:UJ}
\end{figure}

\begin{figure}[ht]
\centering
\includegraphics[clip,width= 12cm]{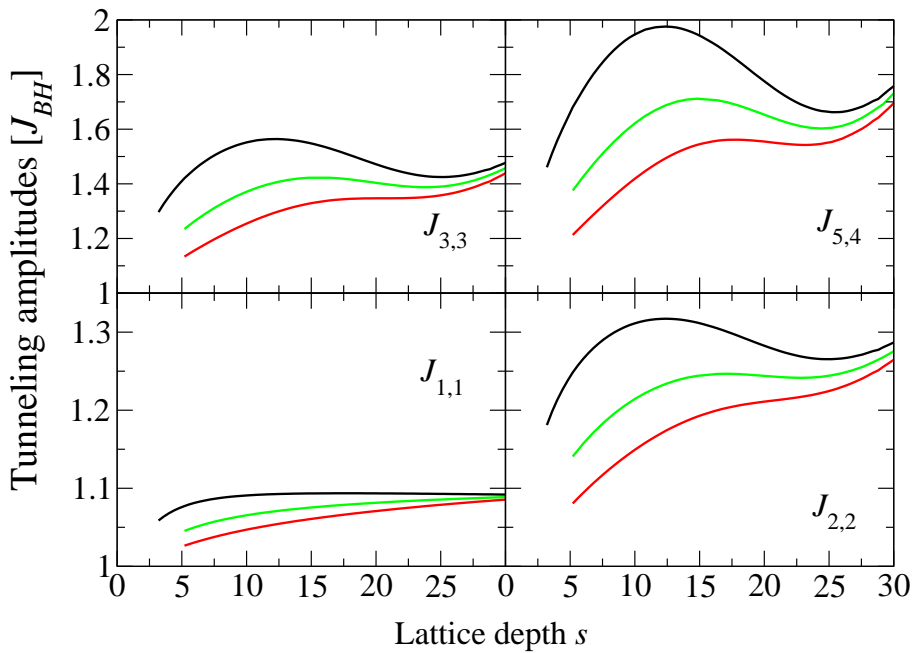}
\caption{Renormalized tunneling amplitudes $J_{n_1,n_2}$ vs. strength of the optical lattice, for various dimensionalities: black, green, and red lines corresponding to 1D, 2D, 3D, respectively. 
Interaction and lattice  parameters: $2a_s/\lambda=0.014,\ \lambda=754
\textrm{ nm},\ s_\perp=34.8.$ The curves meet at
$s=s_\perp.$
} 
\label{fig:Jplots}
\end{figure}

The $U_n$ parameters for the 1D, 2D and 3D lattices  are presented in Figure~\ref{fig:UJ}, while the renormalized 
tunneling amplitudes are shown in Figure~\ref{fig:Jplots}. 
We find that, in low dimensions, $U_n$ vary less with $s$ compared to the full 3D
lattice. The high transverse lattice causes  significant renormalization of $U_n$ even for small $s,$ as $s_\perp$ is still large.

Inspection of Fig.~\ref{fig:comp} and Fig.~\ref{fig:UJ} shows that the difference between consecutive $U_n$ is approximately
constant, i.e.,
\be U_{n-1}-U_n\approx W,
\label{simple}
\ee
 at least for low densities (for typical lattice parameters, $W$ is constant up to 10\%). 
This is  easily understood: 
the alternative effective theory of \cite{Johnson2009} expresses the correction to the on-site energy
term via a three-body interaction term (Eq.~(12) of \cite{Johnson2009}) and $W$ is simply related to their parameter $\tilde U_3$.
The deviations from the linear form, eq.~ (\ref{simple}), may be then related to higher order terms in \cite{Johnson2009}, i.e.
four-body term, etc. Similarly, the lowest order perturbative term discussed above gives a correction to the interaction energy
term $n(n-1)U/2$  of the order $n^3[U^{2000}]^2/\Delta$.  That yields
a crude estimate for  $W\approx U^{2000}/\Delta\approx 1/60,$ in good agreement with Fig.~\ref{fig:comp}.
  
\section{Mean field diagrams for different lattice dimensions}

A  homogeneous (without external trap, $V_i=0$) Hamiltonian may be taken to the thermodynamic limit. 
Then the particle density is determined by the chemical potential $\mu$ and the Hamiltonian parameters: $J$ and $U.$
A MI phase is determined by integer density $\langle\hat{n}\rangle$ and noncompressibility: $\frac{\partial \langle\hat{n}\rangle
}{\partial \mu}=0.$ The rest of the phase diagram is the compressible SF phase
\cite{Fisher1989,Freericks1994,Freericks1996,Kuhner1998}. We perform now the mean field analysis of the phase diagrams of
Hamiltonians (\ref{eqn:HamBH}) and (\ref{eqn:HamEff}). 
A Gutzwiller analysis of the Bose-Hubbard model is a~variational minimization
of the following functional (i.e. mean ground state energy):
\begin{equation}
H^{BH}[\psi]= \langle\psi|H^{BH}|\psi\rangle = -2zJ \langle\hat  a_i \rangle \langle\hat  a_j \rangle + \frac{U}{2}\langle  \hat
n(\hat n-1)\rangle -\mu \langle\hat n\rangle
\label{eqn:FuncBH}
\end{equation}
using the Gutzwiller ansatz: $|\psi\rangle = \bigotimes |\psi_{l}\rangle$ with $|\psi_{l}\rangle= \sum f_n |n\rangle$ the
on-site wavefunction. 
The influence of the lattice geometry is reduced only to the coordinate number, $z$, in the first term, as $\langle\hat a_i \rangle
= \langle\hat a_j \rangle$ due to the translational invariance in thermodynamic limit. 
The phase diagram depends only on the single
parameter $\frac{zJ}{U}.$

\begin{figure}[ht]
\centering
\includegraphics[clip,width= 12cm]{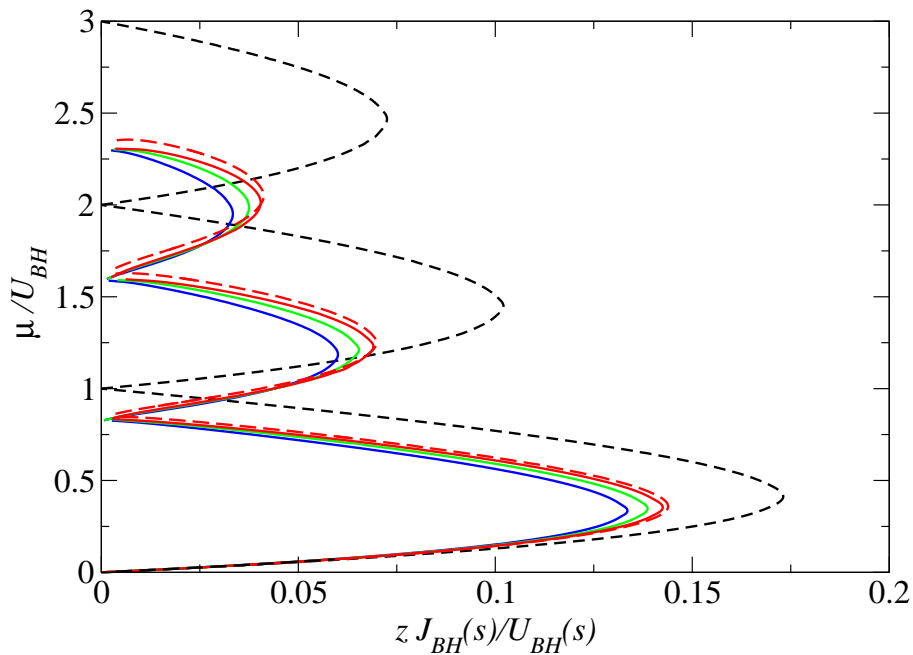}
\caption{Mean field  phase diagrams for 1D, 2D, and 3D lattices. Different curves denote borders between MI and SF phases. 
Dashed black lines correspond to the standard BH model for any dimension, blue, green, and red curves denote 1D, 2D and 3D lattices of
the EMO  Hamiltonian. Dashed red lines show the result obtained for 9 bands as in \cite{Luhmann2012}.
The limit $z J_{BH}(s)/U_{BH}(s), s\to\infty$ is different for each dimension. The $s\to\infty$ limit corresponds to the ill-defined
situation in which the transverse lattice is shallower than the main lattice (this formal limit is also dimension-dependent).
The perpendicular lattice depth is fixed at $s_\perp=34.8$, $\lambda=754 \textrm{ nm}$, $2a_s/\lambda=0.014$
as appropriate for $^{87}$Rb \cite{Luhmann2012}.
}
\label{fig:GutzwAllDim}
\end{figure}

For the EMO Hamiltonian, the dependence of interaction parameters on the dimensionality of the optical lattice is nontrivial. We
shall use the data  from Figure~$\ref{fig:UJ}.$ Thanks to the translational invariance, the~Gutzwiller mean field approach 
to the effective Hamiltonian (\ref{eqn:HamEff}) is equivalent to the minimization of the following functional:

\begin{eqnarray}
H^{EMO}[\psi]&=&  -2z \sum\limits_{n_1,n_2}J_{n_1,n_2} \langle \psi_l  | a_i | n_1 \rangle \langle n_1 | \psi_l \rangle \langle
\psi_l  | a_i^\dagger | n_2 \rangle \langle n_2| \psi_l \rangle +\nonumber \\
 &&+\sum\limits_n   \frac{U_n}{2} n(n-1)|\langle \psi_l |  n \rangle  | ^2 - \mu \sum\limits_n n|\langle \psi_l| n \rangle|^2 .
\label{eqn:FuncefBH}
\end{eqnarray}
 Clearly a~single parameter is no longer sufficient to describe the mean field problem. 
Let us denote by $J_{BH}$ the tunneling amplitude and by $U_{BH}$ the interaction of the standard BH hamiltonian.
 As $\frac{zJ_{BH}(s)}{U_{BH}(s)}$ is strictly decreasing  with the lattice height $s$, it provides  a~way to plot
results calculated for lattice depth $s$ in the $(\frac{zJ_{BH}}{U_{BH}},\mu)$ coordinate space used for a~traditional phase
diagram. This mapping allows also  to directly compare the results obtained using Eqs.~(\ref{eqn:FuncBH}) and
(\ref{eqn:FuncefBH}). Figure~\ref{fig:GutzwAllDim} shows the Gutzwiller phase diagrams for 1D, 2D, 3D lattices. In contrast
with the ordinary BH model, there is a nontrivial dependence on the dimension.
It is rather small for the first lobe becoming more significant for higher occupation of sites. 
Let us stress again that phase diagrams for the MO parameters $J_{n_i,n_j}/U_n$ do not need to be directly related to
$\frac{zJ_{BH}}{U_{BH}}$. The physical observable that is common for the ordinary Bose-Hubbard phase diagram 
and the MO effective
theory is the lattice depth $s.$
Moreover, for dimensions 1 and 2, there is actually {\it a~whole family} of different phase diagrams  indexed by $s_\perp.$ We show
just a~single choice for a~generic value of $s_{\perp}=34.8$.

For 3D, a comparison with the  mean field diagram obtained for 9 bands  \cite{Luhmann2012} is possible. The difference is quite small, the difference in $U_n$ (of the order of few \% - see Fig.~\ref{fig:comp}) manifests itself mostly in shifting the borders between different Mott lobes for higher occupation numbers. Recently a continuous space quantum Monte Carlo calculation of the SF-MI border in the cubic lattice for unit filling has been reported for various scattering lengths \cite{Pilati2012} and compared with the mean field results of \cite{Luhmann2012}. We consider a single ratio of $a_s/\lambda$ as appropriate 
for $^{87}$Rb, the small difference between our results and that of \cite{Luhmann2012} for unit filling  on the scale of Fig.~1 of \cite{Pilati2012} is negligible.   

\begin{figure}[ht]
\centering
\includegraphics[clip,width= 12cm]{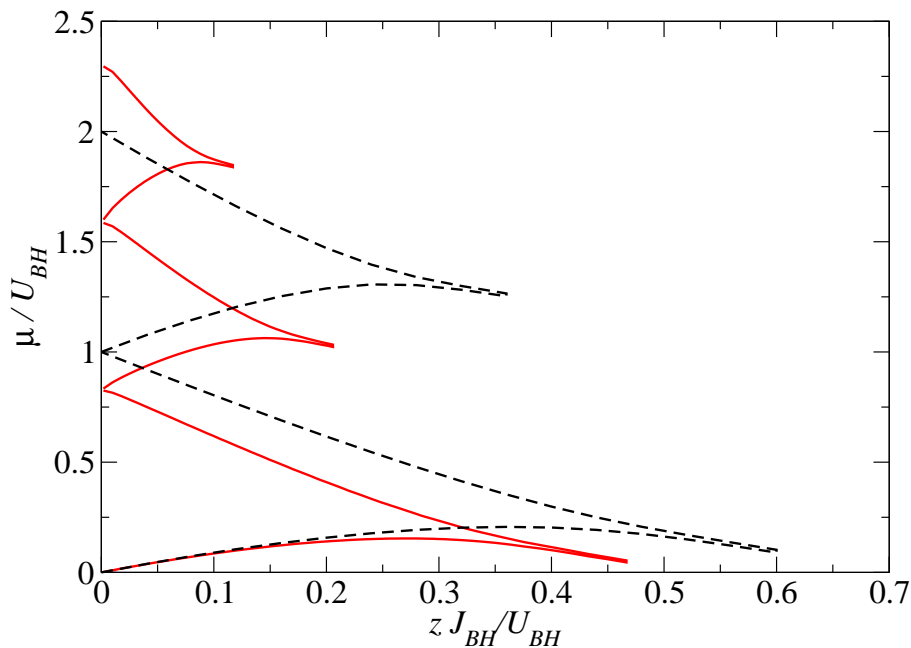}
\caption{1D phase diagram obtained using imaginary time evolution and the TEBD algorithm. 
Black dashed curves presents the standard BH 1D case, the red solid lines are obtained for the EMO 1D model
(\ref{eqn:HamEff}) with $s_\perp=34.8$.
}
\label{fig:PD}
\end{figure}

In 1D, the mean field approximation is  inaccurate. To get a~reliable phase diagram,  we have used energy minimization
through imaginary time evolution using the Time Evolving block Decimation (TEBD) \cite{Vidal2003,Vidal2004} algorithm. 
We fix the lattice size to be $L=100$ (we have checked that choosing a~larger lattice size $L=200,300,400$ does not alter the
results significantly, except at MI tips, where an approximate finite size scaling is performed). This is significantly less computationally
demanding than using the infinite, translationally invariant version of TEBD \cite{Vidal2007,Zakrzewski2008}. 
The transverse lattice height is again $s_\perp=34.8$. Let us denote by $E(N,s)$ the ground state energy of 
a $N$-particle system for lattice
height $s.$  We calculate approximations to the critical values of chemical potentials delimiting a~Mott
insulator region with average filling $n$ by $\mu_+(s)\approx E(nL+1,s)-E(nL,s)$, $\mu_-(s)\approx E(nL,s)-E(nL-1,s).$ We plot
the phase diagrams for both the BH and the EMO Hamiltonians in Figure~\ref{fig:PD}. 
We again see similar results: the Mott lobes
shrink also in 1D, as predicted by the mean field approach. As shown below, this is also
reflected in the dynamical properties.

\section{Consequences of coupling constants renormalization for dynamics}

\subsection{Modulation of optical lattice - absorption spectroscopy}
\label{subsec:modulation}

By periodically modulating the lattice depth, one transfers energy to the atomic sample in the lattice.
Absorption spectroscopy - also incorrectly nicknamed modulation spectroscopy - consists in studying
the dependance of the energy absorption rate with the modulation frequency.  
This absorption is sensitive to the quantum phases present in the system, as shown in early experiments 
\cite{Stoferle2004}.   
It has been simulated \cite{Kollath2006,Huo2011,Lacki2012} for
atoms in an optical lattice in the presence of a~harmonic confinement, using a~standard 1D BH model. 
It seems interesting to
see whether excited bands affect the absorption spectra.  To this end,
we consider the~real time evolution of the ground state of a~given system at
$s=s_0,$ exposed to a 
time-varying lattice height $s(t)=s_0 + s_m \cos\omega t,\ s_m/s_0 \ll 1.$ The simulation is performed in the 
presence of an harmonic trap $V_i=\kappa (i-i_0)^2.$  
The energy of the system is measured after some fixed time.
The energy gain (per particle) as a~function of the modulation frequency
yields the absorption spectrum.

\begin{figure}[ht]
\centering
\includegraphics[clip,width= 12cm]{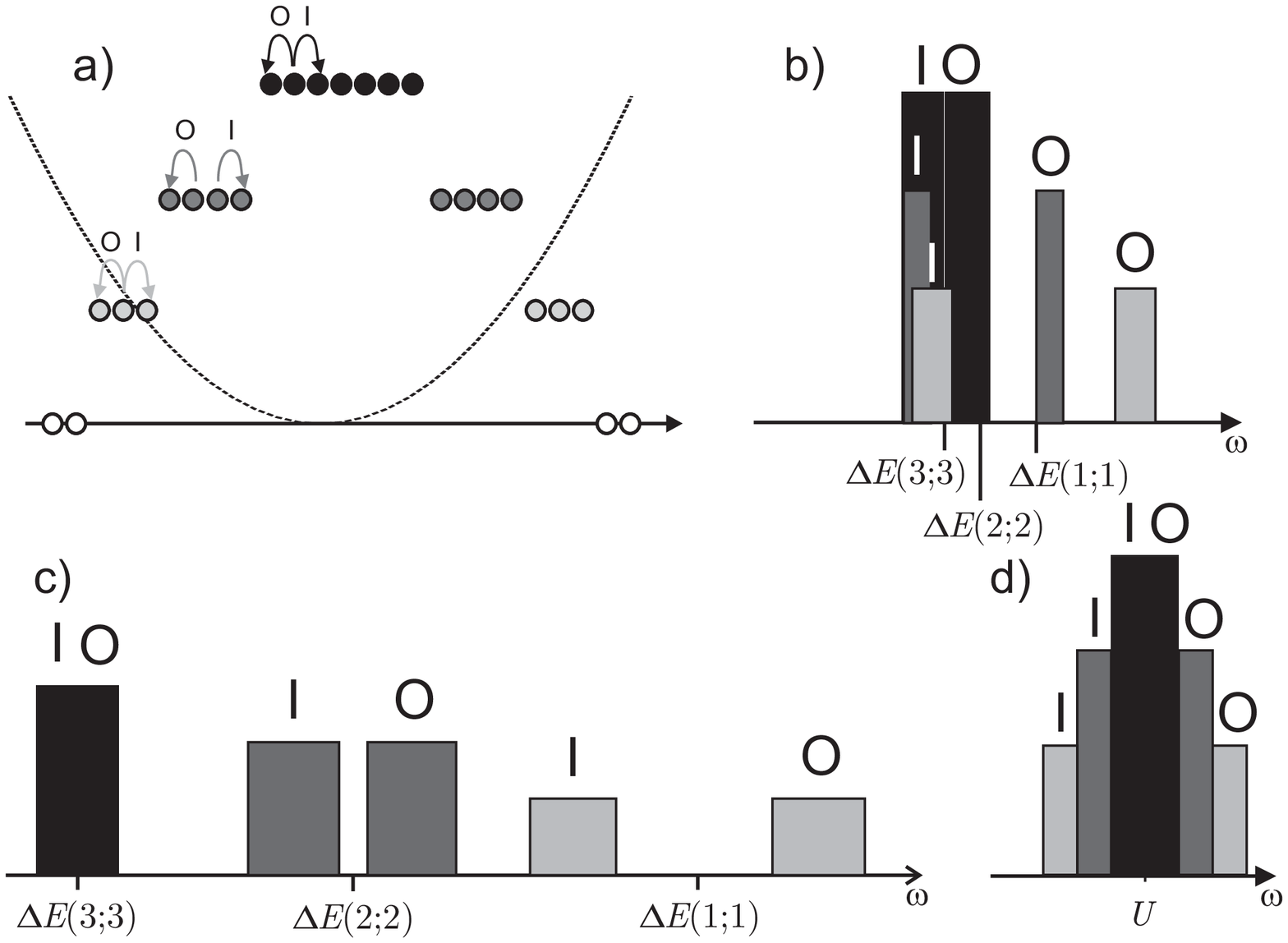}
\caption{
Effects of higher Bloch bands on absorption spectroscopy in the deep Mott (low $J$) regime, $s=15,\ s_\perp=40$. 
Panel (a) shows the~well-known
wedding cake structure with
$n=1, n=2, n=3$ Mott plateaus. Excitations within each plateau 
(colored respectively light gray, dark gray, black, for $n=1,2,3$)
have energies depending on the Mott plateau density and the trapping potential. 
Inward and Outward hopping lead to a splitting
of the absorption structure, a~partial splitting for moderate harmonic trap [(b),  $\kappa=0.001$]
or a~broad well resolved structure for a~shallow trap [(c),  $\kappa=0.0001$] 
in contrast to the standard BH case (d).}
\label{fig:im}
\end{figure}

For a deep optical lattice  in a~predominantly Mott insulating phase, the absorption
spectrum for the standard BH Hamiltonian consists of a~few peaks located at multiplicities of $U$
\cite{Kollath2006,Huo2011,Lacki2012}. The situation is slightly more complex for the EMO Hamiltonian.
The position of peaks can be easily determined in the deep Mott regime ($J\to0)$. 
States excited during the modulation are mainly those that differ from the 
ground state by nearest-neighbours transfer of one particle \cite{Lacki2011,Huo2011}. 
Up to a~small correction due to the difference of the local chemical potential $\mu_i=\mu-V_i,$ 
the excitation energy is determined by the occupation numbers of the source site $i$ and destination site $j.$ It is:
\begin{eqnarray}
\Delta E(n_i;n_j)&=& \frac{1}{2} \left[ -U_{n_j} n_j(n_j-1) + U_{n_j+1} n_j(n_j+1)\right.\nonumber \\
&& \!\!\!\!\!\!\!\!\!\!\!\!\!\!\!\!\!\!\!\!\!\!\!\! \left.-U_{n_i} n_i(n_i-1) +U_{n_i-1} (n_i-1)(n_i-2)  + \Delta \mu_{ij}\right]
\label{eqn:shift}
\end{eqnarray}
with $\Delta\mu_{ij}=\mu_i-\mu_j=V_j-V_i.$ 
Nearest neighbours excitation means that $|i-j|=1,$ and $|n_i-n_j|\leq 1.$
If $U_n=U$ (BH Hamiltonian), we have that $\Delta E(n_i;n_j)=  (n_j-n_i+1) U + \Delta \mu_{ij}.$
By virtue of Eq.~(\ref{simple}) we may approximate Eq.~(\ref{eqn:shift}) by:
$\Delta E(n_i;n_j)=(n_j-n_i+1) U_{n_i} + (n_i-n_j-2)(n_i+n_j-1) W.$ 
For a~trapped gas with maximum occupation number $n=3$, the
relevant values are: $\Delta E(1;1)= U_2, $ $\Delta E(1;2)= 3U_3-U_2\approx 2U_2 - 3W,$ $\Delta E(2;2)= 3U_3-2U_2\approx U_2 - 3W,$ 
$\Delta E(2;3)=6U_4-3U_3-U_2\approx 2U_2 - 9W, $ $\Delta E(3;3)= 6U_4-6U_3+U_2 \approx U_2 - 6W,$ 
$\Delta E(n;n-1)=0,$ with $\Delta\mu $ neglected for clarity. 

A qualitative  comparison of the expected absorption spectra for standard BH case and the EMO model is possible. 
The density profile in the large $s,$ low hopping, limit shows the~well known wedding cake structure 
(see Figure~\ref{fig:im}a). Weak, periodic
modulation leads mainly to nearest neighbour excitations between any pair or neighbouring sites. 
For the standard BH Hamiltonian,
the excitation spectrum consists of a~large peak at energy $U$ (Figure~\ref{fig:im}d). The nonzero width 
of the~peak is due to variations of the local chemical potential (the presence of a~trap): the shallower the trap, the narrower the peak.

For the EMO Hamiltonian, 
the particle-hole excitations from different Mott plateaus have different mean excitation energies. The 
shift with respect to  the mean value is determined by  
the $\Delta\mu_{i,j}.$ For shallow traps  $\Delta\mu_{i,j} \approx 2 (j-i) \kappa(i-i_0)$ and 
the excitation spectrum from  Mott plateau $n$ consists in two bands --- 
one corresponding to Inward (I) hopping, $\Delta\mu_{i,j}
<0$, the other one to Outward (O) hopping,  $\Delta\mu_{i,j} >0$,  with respect to the trap centre.
For the central Mott plateau with density $n_{\textrm{max}}$  a~single, broad peak in the excitation spectrum emerges. 
This is clearly visible in Figures \ref{fig:im}b,c. 
Two cases have been studied: a~system of $N=260$ particles in a trap with $\kappa=0.001$ [(b),
moderate case] and $N=700, \kappa=0.0001$ [(c), very shallow trap]. 
Both cases were studied for the 1D optical lattice with $s=15,$ $s_\perp=40,$ $\lambda=830\textrm{ nm}, a=5.1 \textrm{ nm}.$
The tunneling $J_{n_i,n_j}$ is artificially set to 0 (deep Mott regime).

\begin{figure}[ht]
\centering
\includegraphics[clip,width= 12cm]{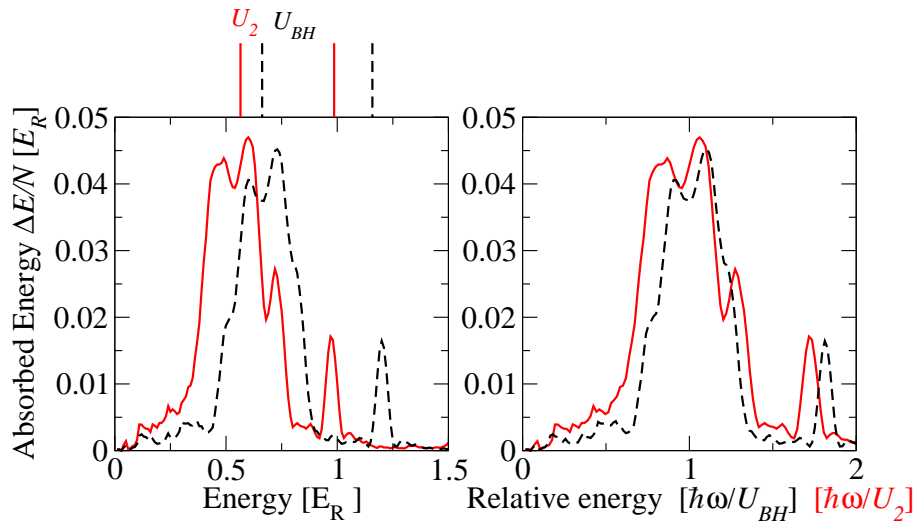}
\caption{Absorption spectrum (modulation time $t=100\hbar/E_R,$ modulation amplitude $s_m=1$). 
Black dashed lines correspond to the BH model,
red solid curves to the effective multiorbital theory.  
Left panel shows spectra on a~common energy scale, observe the
significant shift of the EMO structure toward smaller energies. 
Bars above the plot give the mean expected positions of peaks for the $n$=2 Mott plateau.  
Right panel shows the same data with rescaled energy axes ($U_{BH}$ for the black curve,
$U_2$ for the red one).
 }
\label{fig:absp}
\end{figure}

A similar, but smaller system is analysed in a~subsequent numerical study, for the same parameters taking fully into account the tunneling effects.
We choose a~much tighter trap with curvature, $\kappa=0.009,$ and use true values of 
the hopping constants $J_{n_i,n_j}.$  The
Wannier function calculations give $U_{BH}=0.662E_R$, 
and the renormalization procedure gives $U_2\approx 0.565E_R, W\approx
0.0125E_R.$  

We fill the trap with $N=36$ particles.  This system is similar to the one studied in \cite{Huo2011,Lacki2012}. 
The system states are represented by MPS vectors and evolved using the TEBD algorithm \cite{Vidal2003,Vidal2004}. 
The ground state of the system, being the initial state for the evolution  is calculated using 
an imaginary time evolution with bond dimension $\chi=50.$ The local
Hilbert space assumes maximal filling of $6$ bosons per site.

The density profiles of ground states of the BH and EMO models are practically the same with a
central plateau of 2 particles per site. 
Thus, any change of excitation frequencies can be interpreted as an effect of coupling
constants renormalization.  We have performed the absorption spectroscopy 
simulation for time $t=100\hbar/E_R.$ The modulation
amplitude of the lattice was $s_m =1.$ The results are presented in Figure~\ref{fig:absp}.

The major difference between the spectra obtained for the BH and EMO Hamiltonians 
is a~significant shift of the observed structures.
While for the BH case, the main structure is centered at $U_{BH}$,
it has a similar shape, but centered around $U_2$ in the EMO case.
Because of the steep harmonic trap - thus the large changes in local chemical potentials - 
the structure of the peaks is rather complex. Note the global broadening for the EMO case,
and an additional peak in the main $U_2$ structure, corresponding to the $n=1$ plateau excitations having
an energy larger by roughly $3W$, as discussed above.

A second small peak on the right appears at $E=1.75U_2$ (EMO case)
and $E=1.8U_{BH}$ (BH case). It corresponds to a~particle-hole excitation on the edge 
between the $n=1$ and $n=2$ Mott plateaus as identified in~\cite{Lacki2012}. 
The right panel shows that the spectra becomes quite similar if rescaled by their proper
energy scale, $U_{BH}$ or $U_2.$
  
The absorption spectra are quite sensitive to the details of the system. Taking the same parameters
for a slightly larger number of particles may create situations where the ground states of the BH and EMO 
Hamiltonians significantly
differ. This is then reflected in the absorption spectra. If the density profile contains 
a $n=3$ or higher plateau - then the structure
of peaks may become more complicated, as discussed above. 

\begin{figure}[ht]
\centering
\includegraphics[clip,width= 12cm]{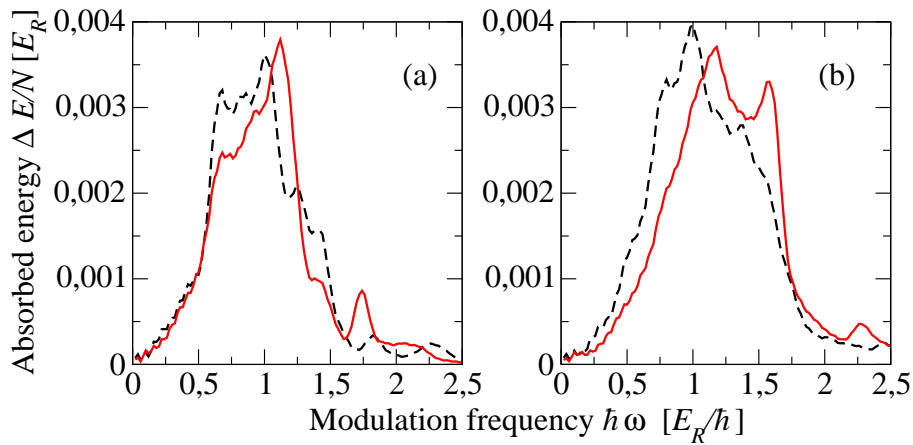}
\caption{Absorption spectra in the superfluid case. The absorbed energy per particle is plotted as a~function of
the frequency of modulation of the lattice depth. 
Here $s=5, s_m=0.2, s_\perp=40, U_{BH}\approx 0.465E_R.$ The renormalized interaction parameter: $
U_2\approx0.406E_R.$  The left panel corresponds to 20 particles, the right panel to 36 particles.}
\label{fig:figsf}
\end{figure}

We have also compared the absorption spectra in the superfluid regime.  
The lattice height is fixed at $s=5,
s_\perp=40$~\cite{Fallani2007}. 
The system is modulated  for $t=50 \hbar/E_R$ with $s_m=0.2$. 
The results are presented in Figure~\ref{fig:figsf}. Unlike in the Mott regime, the
positions of the absorption peaks are not determined solely by the interaction. 
In particular, no global shift of the structure is observed. In both cases, one observes a~broad resonance 
around the recoil energy, with complicated detailed structures.   
Note that the modulation depth is much smaller than in the Mott insulator situation
(to avoid significant excitation of the system) and therefore 
the absorbed energy per particle is much smaller than in
Fig.~\ref{fig:absp}.

\subsection{Florence experiment revisited}

In the Florence experiment~\cite{Fallani2007}, the starting point is a ultra-cold gas in a~harmonic trap (without
optical lattice). The optical lattices are then ramped up
(assuming $s(t) =0.2 s_\perp(t)$) with an exponential ramp $s(t) \sim s_0(1-\exp(t/\tau)),$ for $\tau=30 $ ms, 
and a~total ramping
duration 100 ms. The system soon becomes quasi-1D producing a~set of 1D tubes. 
If  a~``disordered'' system is desired, an
additional optical lattice, with different wavelength $\lambda_2$, is superimposed along the tubes. 
This adds a
potential $V_d(x)=s_2 \sin^2(k_2 x).$  For $s_2\ll s$, it acts effectively as the shift of  
the on-site energy, i.e.
an additional pseudo-random disorder $\Delta V_i = s_2 \sin^2\left(\frac{\lambda}{\lambda_2} \pi i + \phi\right),$ 
where $\phi$ represents the offset between the two optical lattices.  
We take a~generic value $\phi=0.12345$. 
If the ratio $\lambda / \lambda_2$ is chosen irrational enough, 
the bichromatic lattice simulates a disorder well enough for
a~ finite system \cite{Diener2001,Damski2003,Roth2003,Roscilde2008,Roux2008}. 
The  dependence of the pseudo-disorder amplitude $s_2$ on
time is set by demanding that $s_2(t)\sim s(t)$
(all optical lattices are ramped up simultaneously). We will consider 3 cases: no disorder ($s_2(t)=0),$ weak disorder
($s_2=\frac{1}{32} s$), and strong disorder ($s_2=\frac{5}{32} s$).

Consider first the no disorder case $s_2=0$. 
As initial state, we take 151 particles on 81 sites 
in the presence of a harmonic confinement coming both from the trap and the transverse laser profile. 
The detailed procedure using the TEBD algorithm is described in \cite{Zakrzewski2009,Lacki2012}. 

After the optical lattice is ramped, absorption spectroscopy is performed
for 30ms (the conversion unit is  $20.91 \hbar / E_R=1\textrm{ms}$). In the recent numerical investigation of this experiment
\cite{Zakrzewski2009}, a~discrepancy between experimental~\cite{Fallani2007} and numerical results
was found. The reported position of the first absorption peak was 1.9 kHz \cite{Fallani2007}, while Wannier
function calculations gave 2.3kHz \cite{Zakrzewski2009}. The renormalization procedure 
renormalizes the value of the $U$ parameter to $U_2=2$ kHz,
$U_3=1.96$ kHz, $U_4=1.91$ kHz.  This suggests that the positions of absorption peaks due to 
the $n=1,2,3$ Mott plateaus are: $U_2=2 \textrm{kHz}$, $-2 U_2 + 3U_3=1.85 \textrm{kHz}$, $U_2-6U_3+6U_4=1.74 \textrm{kHz}.$ 
The "average" peak position is 1.87 kHz. Therefore, the EMO Hamiltonian 
provides an estimate of the peak position in good agreement with the
experiment. 

\begin{figure}[ht]
\centering
\includegraphics[clip,width= 12cm]{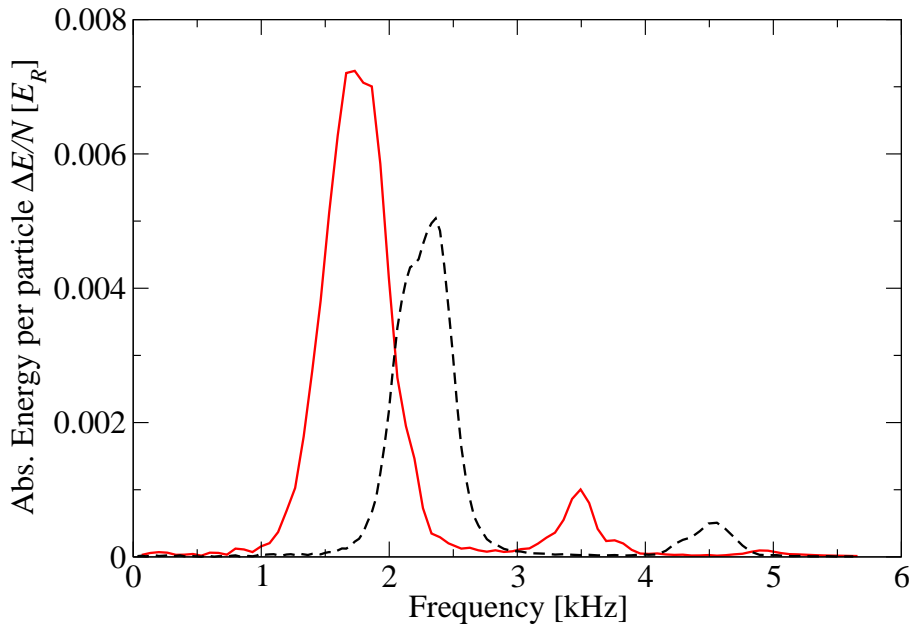}
\caption{Absorption spectrum obtained by applying lattice modulation with amplitude $s_m=1$ 
on the wavepacket created by exponential ramp up to $s=16$. 
The black dashed line corresponds to the standard BH model, the red line is the result
of effective multiorbital theory. The position of the absorption peaks in the latter case reproduce well 
 the experimental results \cite{Fallani2007}.} 
\label{fig:AbsIngu}
\end{figure}

A~simulation of absorption spectrum  performed for this system confirms this finding as shown in 
Figure~\ref{fig:AbsIngu}. Although the initial state when the periodic lattice modulation
starts is not the ground state, but a wavepacket dynamically created during the ramping of the
lattice, the peak positions are well predicted by the EMO model.
The position of the first and second peaks agree quite well with the experiment (the relative height is
different presumably because of the strong modulation used in \cite{Fallani2007}).

The exponential ramping of the optical lattice in the experiment \cite{Fallani2007} may not be adiabatic as discussed  in
\cite{Zakrzewski2009} using a~standard BH description. 
Due to the discrepancy in the position of the absorption peak, the lattice depth was adjusted 
in \cite{Zakrzewski2009}. Instead of ramping the lattice up to
$s=16,$ the final value $s=14$ was considered. In some sense, 
such a~simplified approach may be viewed as a renormalization of the BH parameters
(without insight into its origin explained in section \ref{subsec:modulation}). 
Let us stress
that the agreement between the experimental position of absorption peaks and the EMO predictions
prove the necessity of using the effective multiorbital theory to explain quantitatively the experimental results. 

With that modification of the final $s$ value,  it was found using the BH model \cite{Zakrzewski2009} 
that the overlap of the prepared wavepacket on the ground state at the final $s$ value was about 
9\% in the absence of disorder. It is most interesting to see how taking
into account higher bands within effective multiorbital theory affects the adiabaticity of the dynamics.  
The simulation performed for the similar exponential ramp starting at $s=5$ up to $s=16$  yields  
an overlap of the dynamical wavepacket on the ground state at
$s=16$ equal to 17.3\%. It may be qualitatively understood: in the EMO model,
the effective interactions are weaker and the effective tunneling larger allowing particles to redistribute more efficiently among sites during the ramp. 

We have also tested an optimized $s(t)$ pulse shape as in \cite{Zakrzewski2009}. By choosing $s(t)$ 
changing slowly close to the phase transition point for the $n=3$ Mott lobe, we have been able to enhance adiabaticity up to 
33\% squared overlap with the ground state.

The presence of disorder has devastating influence on adiabaticity, similarly to the standard BH case \cite{Zakrzewski2009}. We have
found that for
a small disorder, $s_2=\frac{1}{32}s$ the squared overlap is a~fraction of percent (0.005) 
while for the strong disorder  $s_2=\frac{5}{32}s$ it becomes vanishingly small (of the order
of $10^{-9}$, beyond the accuracy of the calculation). 

\section{Conclusion}

The aim of this paper is two-fold. In the first part, we have presented an efficient numerical implementation 
of the approach sketched in \cite{Luhmann2012} which makes it possible to compute 
the parameters of the effective Hamiltonian for bosons in optical
lattices. 
The approach goes beyond the standard Bose-Hubbard model \cite{Jaksch1998} limited to the lowest Bloch band. The
effective Hamiltonian approach which includes contributions from higher lying bands (multiorbital approach)  
has been shown to lead
to new effects and even new phases \cite{Dutta2011} for bosonic systems (see also \cite{Johnson2009}). Our  scheme of perturbatively
generated basis seems clearly superior to the energy-selected basis used in~\cite{Luhmann2012} 
for low and moderate occupation numbers 
and allows for better estimates of Hamiltonian parameters. 
These estimates may be extrapolated to an infinite number of
bands.

We have applied the  method  not only to the standard 3D cubic lattice, but also to reduced 1D and 2D problems, where the 
lattice depth is different in various directions. 
The effective Hamiltonian obtained depends on the dimensionality of the problem.  In effect,
mean field phase diagrams as obtained with the Gutzwiller ansatz, differ even if they are rescaled by the lattice coordination
number.  It turns out that the role of excited bands is even more pronounced for reduced dimensionality problems than for a~3D
lattice.

Motivated by this difference, we have investigated whether the dynamics is different in a~standard Bose-Hubbard model and for the
effective multiorbital theory. 
We have considered two cases, the energy absorption created by modulation of the lattice height and the
quasi-adiabatic passage from the superfluid to the Mott insulator phase. 
In both situations, it turns out that taking
into account the density dependent tunneling terms as well as modification of interactions
may lead to significant differences between two approaches.
For the same lattice depth, the effective interactions turn out to be
significantly weaker than in the standard Bose-Hubbard model. 
This results in profound differences in the absorption spectra such as
significant shifts of absorption peaks. 
Similarly, the full effective theory predicts that the transition from
superfluid to Mott insulator is more adiabatic than with the standard
Bose-Hubbard model \cite{Zakrzewski2009}.  

The results presented in the present paper should have a~direct applicability to any 
experiment using bosons in an optical lattice, with multiple site occupations.

\ack
We thank an anonymous referee for pointing out the potential problem
of contact interaction potentials in 3D calculations.
M.\L{}. acknowledges communications with D.-S. L\"uhmann on the details of calculations in \cite{Luhmann2012}. This work was supported
by the International PhD Projects Programme of the
Foundation for Polish Science within the European Regional Development Fund of the
European Union, agreement no. MPD/2009/6. J.Z. acknowledges partial support from Polish National Center
for Science grant No. DEC-2012/04/A/ST2/00088.

\appendix
\section{Diagonalization of the onside Hamiltonian}

We describe  the approach we use to generate a perturbatively based variational set used in the diagonalization of the onsite $n$ particle problem:
\begin{equation}
\mathcal{H}_{loc}= H_E + H_U = \sum\limits_\alpha E_\alpha \hat n_\alpha + \sum\limits_{\alpha\beta\gamma\delta} U^{\alpha\beta\gamma\delta} \hat a^\dagger_\alpha  \hat a^\dagger_\beta  \hat a_\gamma  \hat a_\delta.
\label{eq:a}
\end{equation}

The most elementary possible excitation promotes a boson from the $\alpha=0$ to the $\alpha=2$ band. This defines two  limits: $n^{3/2}U^{2000} \ll E_2-E_0$ and $n^{3/2}U^{2000} \gg E_2-E_0.$ $U^{\alpha\beta\gamma\delta}$ become smaller as $\alpha,\beta,\gamma,\delta$ indices grow.
 
The first regime corresponds to the ordinary Bose-Hubbard model in which $H_U$ can be treated as a
small perturbation, with zero order ground state $|\psi_0\rangle=\left(\hat a_{(0,0,0)}^\dagger \right)^n| 0 \rangle,$ just as assumed for the ordinary BH hamiltonian. The large density regime is dominated by $H_U.$ A straightforward application of the multiband model in that regime is not justified, however a~mean field based renormalization scheme leads to the approximate model of the same form with modified parameters \cite{Oosten2003}. The transition to the nonperturbative regime occurs at approximately 10-15 atoms per site.

\begin{figure}[ht]
\centering
\includegraphics[clip,width= 12cm]{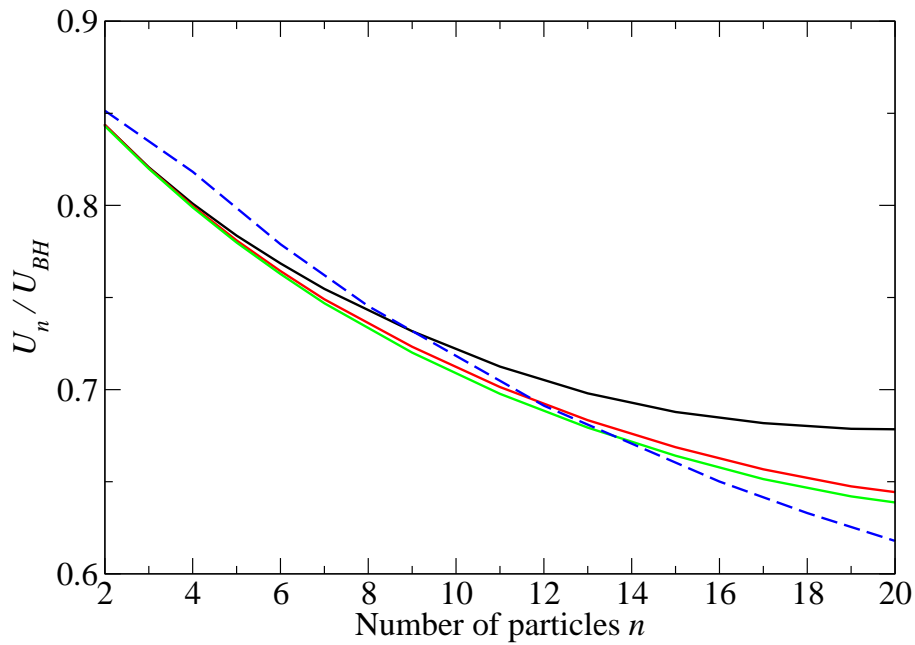}
\caption{Comparison of $U_n$ obtained from diagonalization of $\mathcal{H}_{loc}$, Eq.(\ref{eq:a}), with energy selected basis  and with perturbatively chosen sets.  Each  perturbative set consists of 10000 vectors of first order and of \ 0, 100, 800 vectors of second order (curves: black, red, green). Blue dashed curve denotes the least-energy basis result. For small and moderate $n$, the perturbative based basis is clearly superior over the least-energy set. The failure of perturbation theory approach for $n$ sufficiently large is apparent too.  }
\label{fig:zebr}
\end{figure}

Consider the low energy regime. By vectors reachable in $k$-th order perturbative expansion, we call those Fock states $|\psi_p\rangle$ for which $\langle \psi_0 | H_U^k | \psi_p \rangle \neq 0.$ In particular a~full basis can be generated with order $\left\lceil\frac{n}{2}\right\rceil$. Let us denote by $\mathcal{B}_k$ the set of vectors reachable in $k$-th order and unreachable in $k-1$-th order. The full variational basis of Fock states is $\mathcal{B}=\bigcup\limits_k \mathcal{B}_k,$ with $\mathcal{B}_k$ pairwise disjoint, and $\mathcal{B}_k = \emptyset, \textrm{ for } k>\left\lceil\frac{n}{2}\right\rceil.$

For numerical diagonalization, a~proper, not too large $\mathcal{S}\subset \mathcal{B}$  has to be chosen. In \cite{Luhmann2012}, $\mathcal{S}$ consists of vectors $|\psi\rangle$ with the least $E_{|\psi\rangle}=\langle \psi | \mathcal{H}_{loc} | \psi \rangle.$ In the first order perturbation method, we choose vectors from   $\mathcal{B}_1$ with the largest values of:

\begin{equation}
f_1( \psi) =  \ln \frac{ |\langle \psi_0 | \hat H_U | \psi \rangle |}{E_{|\psi\rangle} -E_{|\psi_0\rangle}}.
\label{eqn:first}
\end{equation}
Disregarding vectors from $\bigcup\limits_{k\ge 2} \mathcal{B}_k$ seems to be a~crude approach, but still our numerical calculations prove this "basic" perturbative approach to be more efficient  for low densities  than the least-energy-based selection. The perturbation theory provides a~way to evaluate a~perturbative contribution of vectors from $\mathcal{B}_k,$ for arbitrary $k,$ which, unfortunately, is computationally involved (summation over intermediate states, degeneracy resolution). As we do not calculate the ground state energy within the perturbation theory treatment, but only motivate the choice of a variational basis for numerical diagonalization, the following function:

\begin{equation}
 f_2( \psi) =  \ln \sup_{|\psi_1\rangle\in\mathcal{B}_1} \frac{ |\langle \psi_0 | \hat H_U | \psi_1 \rangle \langle \psi_1 | \hat H_U | \psi \rangle | }{(E_{|\psi\rangle}-E_{|\psi_0\rangle})(E_{|\psi_1\rangle}-E_{|\psi_0\rangle})}.
\label{eqn:second}
\end{equation}
 is chosen to approximate the relevancy measure for vector $|\psi\rangle.$ It minimizes the state's $|\psi\rangle$ energy and maximizes the overlap.

Detailed comparison of the three methods: the least-energy, first order perturbative and "improved" first order perturbative approach is presented in Figure~\ref{fig:comp}. We have used $K=10000$ vectors (plus  additional second higher order vectors for the "improved" method), set $s=s_\perp=38$ and compared the three methods as a~function of $n.$ Two regimes: perturbative and not perturbative emerge, as expected. In the small and moderate $n$ regime, the perturbative approach gives better estimate for the ground state energy than  the least-energy method \cite{Luhmann2012}. The opposite tendency is visible in   the nonperturbative regime.
 
Let us describe in detail how a choice of the basis with the largest $f_1$ and $f_2$ values is performed. We have fixed the maximal number of Bloch bands included at $B=15.$ We use Markov chain Monte Carlo method which is quite general and can be applied at any order $k$ of perturbation expansion. $\mathcal{B}_k$ is in general too large to evaluate function $f_k$ for all elements (its size increases exponentially with the total number of particles).  It is usually possible just to scan the whole $\mathcal{B}_1$ set, so from this point on we assume $k\geq 2$. To choose $K$ vectors with largest $f_k$ values, we construct a~random walk based on Metropolis' Monte Carlo algorithm. 
A state of the random walk is a~finite $k+1$-tuple of $n$-particle Fock states $\mathcal{V}=(|\psi_0\rangle,|\psi_1\rangle,|\psi_2\rangle,\ldots,|\psi_{k_1}\rangle,|\psi\rangle),$ for $|\psi_i\rangle\in \mathcal{B}_i.$  For any such $\mathcal{V}$ we define the following generalization of (\ref{eqn:second}):

\begin{eqnarray}
 g_k(\mathcal{V})=  \ln \frac{ |\langle \psi_0 | \hat H_U | \psi_1 \rangle \langle \psi_1 | \hat H_U | \psi_2 \rangle |\ldots \langle \psi_{k-1} | \hat H_U | \psi \rangle |}{(E_{|\psi\rangle} -E_{|\psi_0\rangle})(E_{|\psi_{k-1}\rangle} -E_{|\psi_0\rangle})\ldots (E_{|\psi_1\rangle} -E_{{|\psi_0\rangle}})},\nonumber \\
 f_k(\psi)= \sup\limits_{|\psi_1\rangle \in \mathcal{B}_1,\ldots,|\psi_{k-1}\rangle \in \mathcal{B}_{k-1} } g_k({\mathcal{V}}).
\end{eqnarray}

To get the random walk, we have to update $\mathcal{V}.$ First we choose at random a~Fock state  $|\psi_l\rangle\in\mathcal{V}$ to be updated. With equal probability, we update one or two particles of $|\psi_l\rangle$ preserving the total parity of the state. One particle update is done according  to $|\psi_l\rangle\to \hat a_{(i_x,i_y,i_z)} \hat a^\dagger_{(i_x,i_y',i_z)}|\psi_l\rangle,  i_y\equiv i_y' (\textrm{mod }2),$ while two particle-update is: 
$|\psi_l\rangle\to \hat a_{(i_x,i_y,i_z)}\hat a_{(j_x,j_y,j_z)} \hat a^\dagger_{(i_x,i_y',i_z)}\hat a^\dagger_{(j_x,j_y',j_z)} |\psi_l\rangle ,  i_y+j_y\equiv i_y'+j_y' (\textrm{mod }2).$  All vectors are normalized. Direction $y$ is not special in any way: with equal probability any of $x,y,z$ is chosen. After the update a~proposition $\mathcal{V}'$ is prepared. We automatically reject updates for which $|\psi_l\rangle \not\in \mathcal{B}_l.$ If that is not the case, the acceptance probability is determined as in Metropolis algorithm: it is given by $\min\{1,\exp[\beta(g_k(\mathcal{V}')-g_k(\mathcal{V}))]\}.$ The inverse temperature $\beta$ is tuned to optimize sampling efficiency --- we choose it by requiring the acceptance rate to be close to 0.3.  After a successful update, the last element of the tuple $\mathcal{V}',$ state $|\psi_k\rangle$ is accepted into the solution set if its perturbative importance $g_k(\mathcal{V})$ is in the $K$ lowest values recorded so far. Th!
 e accepted vector $|\psi_k\rangle$ is memorized as well as the importance value $g_k(\mathcal{V}').$ If $|\psi_k\rangle$ had been generated before, the memorized value of $g_k$ is updated (only if the new value of larger than the old one). If, in a~subsequent few thousand sweeps (empirical value), no vector makes it into the solution set, nor $g_k$ values are updated, then the procedure is restarted. This ensures that all low energy excitations are taken into account (the starting point is always the low energy configuration).  Altogether, we make $2\times 10^9$ MC sweeps to generate basis of size 40000 (as used for the results presented in the main part of the paper).

If all Bloch bands were included, then the set of Fock space would be infinite. On the other hand, only a finite number of them could satisfy the inequality: $ f_k > \varepsilon.$ The values of $f_i$ for the remaining states are very close to 0, and a singularity in density of states $\frac{\partial f_k}{\partial \vec n}$ arises. Logarithm is used to "smoothen" this singularity for numerical purposes. It does not affect the ordering, as $\ln$ is increasing injection. 


\section*{References}

\providecommand{\newblock}{}


\end{document}